\documentclass{jltp}

\usepackage{amsfonts,amsmath,graphicx,latexsym}
\usepackage{epsfig}

\newcommand{\vc}[1]{\mathbf{#1}}

\title{Kinetic theory and dynamic structure factor of a condensate  
in the random phase approximation}
\author{Patrick Navez}

\address{
Ecole Polytechnique, CP 165, Universit\'e Libre de Bruxelles, 1050
Brussels, Belgium}

\runninghead{P. Navez}{Kinetic theory and dyn. struct. factor of a condensate in the RPA}

\begin{document}
\maketitle

\begin{abstract}
We present the microscopic kinetic theory of a 
homogeneous dilute Bose condensed gas in the generalized random phase 
approximation (GRPA), which satisfies the following requirements:
1) the mass, momentum and energy conservation laws; 2) 
the H-theorem; 3) the superfluidity property and 4) 
the recovery of the Bogoliubov theory at zero temperature 
\cite{condenson}. 
In this approach, the condensate influences 
the binary collisional process between the two normal atoms, in the sense that 
their interaction force results from the mediation of a Bogoliubov 
collective excitation traveling throughout the condensate.
Furthermore, as long as the Bose gas is stable, no collision happens 
between condensed and normal atoms. In this paper, we show how the 
kinetic theory in the GRPA allows to calculate the dynamic structure 
factor at finite temperature and when the normal and superfluid are 
in a relative motion. The obtained spectrum for this factor provides a prediction 
which, compared to the experimental results, allows to validate 
the GRPA.

\end{abstract}

\section{INTRODUCTION}
The kinetic theory of an interacting diluted Bose condensed gas remains still 
a challenging problem. The situation is indeed not entirely clear 
even though some microscopic description of the condensate 
dynamics have been attempted by many groups \cite{Leggett,Zaremba,Walser}. 
Some of the difficulties are really related to the basics of kinetics 
that Boltzmann developed more than one century ago, in order 
to derive an equation for the particle position
and momentum distribution in a classical gas. 
This equation obeys the conservation laws for the mass, 
the momentum and the total energy density as well as the famous 
H-theorem which allows to recover the second principle of 
thermodynamics. Strangely,
the extension of that approach 
to a quantum Bose condensed gas compatible with these 
conservation laws and the H-theorem has not seemed to attract a 
lot of attention.
In addition, the kinetic equation must describe 
the superfluidity, in particular, the persistence of 
a non zero relative velocity between the normal and superfluid 
when it is below the critical velocity.
Also, we expect to recover at zero temperature the 
Bogoliubov theory whose validity has been confirmed experimentally 
\cite{Ketterle}.
At first glance, all these requirements seem logical and 
well defined but an analysis of the literature 
reveals the difficulties to satisfy them in a unified theory.
But, a solution to this problem has been recently 
presented 
with surprising properties
\cite{condenson}.

A kinetic equation has been derived by extending the generalized 
random phase approximation (GRPA),  generally used for describing 
charged particles, to the case of a condensed atom gas. 
For a system of charged particles, the resulting equations describe 
the binary collision between the particles  as a process of emission 
and absorption of  a plasmon mediating the interaction. This 
plasmon corresponds to the longitudinal component of an electromagnetic wave 
in addition to the transverse ones. For neutral bosons, however, 
the condensate itself helps the thermal atoms to collide  by transferring 
momentum and energy. The vehicle for this transmission is 
the Bogoliubov collective excitation energy with its dispersion relation 
characterizing a Goldstone boson (see Fig.1). 
\begin{figure}
\centerline{\scalebox{0.8}{
\includegraphics{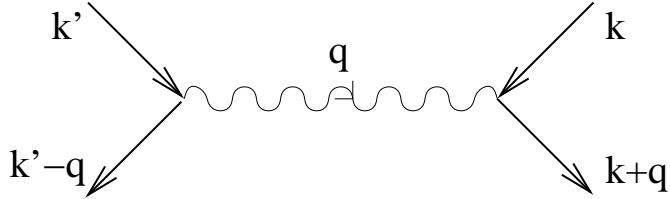}}}
\caption{Feymann diagram of the
interaction of two normal atoms (full line)
mediated by the Bogoliubov collective excitation (wavy line).}
\end{figure}
Surprisingly also, provided that the condensate is stable, 
particle exchanges  
between the condensate and the thermal cloud are excluded since no collisions 
allow such a process. 
This point can be understood qualitatively by analogy with salt in the water. 
The interaction force between ions of salt is 
strongly attenuated due to the dielectric power of water to generate 
a counter polarized force. In the same way, 
due to its powerful ``dielectric'' behavior, 
the condensate can deform its wave function in a coherent 
way in order to diminish its potential energy  and attenuate 
locally the potential interaction force displayed by any excited atom.
In the GRPA, this effect is so strong that any atom in the condensate 
does not feel the presence of an excited atom anymore leading to the 
absence of any 
collision process. 
As a result, the particles of the normal and superfluid become 
invisible to each other and feel only the mean field force of the other 
fluid. This different picture allows to understand more simply why a relative 
velocity between the two fluids persists forever and leads to a metastable 
state characteristic of the phenomenon of superfluidity. 
This metastable state persists in time provided that the relative 
velocity does not exceed the sound velocity. Otherwise, the Landau 
criterion is not satisfied and the gas becomes unstable leading 
to some particle exchanges again. 


The purpose of this paper is to show how to calculate the dynamic  
structure factor in the GRPA at finite temperature and in presence 
of a relative velocity. Since this factor can be measured 
in experiment, comparisons can serve to validate the GRPA approach. 
After recalling the basics of 
the GRPA method for the kinetics of a Bose gas in section 2, 
we show how to calculate this factor in section 3.

\section{THE KINETIC EQUATION IN A HOMOGENEOUS GAS}

For any momentum $\vc{q}$, 
we define the density operator $\rho_{\vc{q}}=\sum_{\vc{k}}
\rho_{\vc{k},\vc{q}}$ where 
$\rho_{\vc{k},\vc{q}}=c^\dagger_{\vc{k}}c_{\vc{k}+\vc{q}}$ and 
$c_{\vc{k}}$ ($c^\dagger_{\vc{k}}$) is the destruction (creation) 
operator.
The GRPA assumes that the diagonal contribution is the dominant one 
i.e. $\langle \rho_{\vc{k},\vc{q}} \rangle = \delta_{\vc{q},\vc{0}}n_{\vc{k}}$.
For 
$\vc{q}=0$, the time evolution in a volume $V$ of the 
population average  
of the atom with mass $m$ in the mode $\vc{k}$ 
is given by \cite{condenson}: 
\begin{eqnarray}\label{n}
i\frac{\partial}{\partial t}n_{\vc{k}}
=\sum_{\vc{q'}}\frac{2\pi a}{mV}
\langle (\rho_{\vc{k},\vc{q'}}
-\rho_{\vc{k}-\vc{q'},\vc{q'}})
\rho^\dagger_{\vc{q'}}+\rho^\dagger_{\vc{q'}}(\rho_{\vc{k},\vc{q'}}
-\rho_{\vc{k}-\vc{q'},\vc{q'}})\rangle
\end{eqnarray}
where $a$ is the scattering length.
For $\vc{q}\not=0$ in the GRPA, we neglect 
contribution that are quadratic in the non-diagonal 
operator in order to get an approximated linear equation:
\begin{eqnarray}\label{rho}
\lefteqn{[i\frac{\partial}{\partial t}
+\epsilon_\vc{k} - \epsilon_{\vc{k}+\vc{q}}]
\rho_{\vc{k},\vc{q}}}
\nonumber \\
&=\displaystyle  \frac{4\pi a}{mV} 
\sum_{\vc{k'}}\big[n_{\vc{k}}
(2-\delta_{\vc{k},\vc{k'}}-\delta_{\vc{k}-\vc{q},\vc{k'}})
-n_{\vc{k+q}}
(2-\delta_{\vc{k},\vc{k'}}-\delta_{\vc{k}+\vc{q},\vc{k'}})
\big]\rho_{\vc{k'},\vc{q}}
\end{eqnarray}
where $\epsilon_\vc{k}=\vc{k}^2/2m$. 
This equation possesses 
two types of eigenvalues \cite{NP}: 1) the scattering 
solution with the spectrum given by the kinetic energy 
transfer: 
$\omega= \epsilon_{\vc{k}+\vc{q}}-\epsilon_{\vc{k}}$; 2)
for excitations involving the macroscopic condensate mode $n_{\vc{k_s}}$ i.e. 
for 
$\vc{k}= \vc{k_s}, \vc{k_s}-\vc{q}$, the presence of
interaction transforms
the scattering solutions
into  collective solutions $\omega=\omega_{\vc{q}}-i\gamma_{\vc{q}}$. 
These are the zeroes 
of 
the dynamic dielectric function (DDF):
\begin{eqnarray}\label{kal}
{\cal K}(\vc{q},\omega)={\cal K}_n(\vc{q},\omega)
-
\frac{\frac{8\pi a n_{\vc{k_s}}}{mV}
\frac{\vc{q}^2}{m}}
{(\omega+i0_+ -
\frac{\vc{k_s}.\vc{q}}{m})^2-(\frac{\vc{q}^2}{2m})^2+
\frac{4\pi a n_{\vc{k_s}}}{mV}
\frac{\vc{q}^2}{m}}
\end{eqnarray}
where $
\epsilon^B_{\vc{q}}=
\sqrt{c^2 \vc{q}^2 +
(\vc{q}^2/2m)^2}$
is the Bogoliubov excitation energy,
$c=\sqrt{{4\pi a n_{\vc{k_s}}}/{m^2 V}}$
is the sound velocity
and
\begin{eqnarray}\label{kn}
{\cal K}_n(\vc{q},\omega)=
1- \frac{8\pi a }{mV}\sum_{\vc{k}}
\frac{n'_{\vc{k}}-n'_{\vc{k+q}}}{
\omega+i0_+ -
\frac{\vc{k}.\vc{q}}{m}-\frac{\vc{q}^2}{2m}}
\end{eqnarray}
is the DDF of the normal
fluid where $n'_{\vc{k}}=(1-\delta_{\vc{k},\vc{k_s}})n_{\vc{k}}$.
The real 
part and imaginary part correspond to 
the Bogoliubov collective excitation and the Landau 
damping respectively. For a temperature close to zero, 
there are two solutions\cite{condenson}:
$\omega^\pm_{\vc{q}} \simeq \frac{\vc{k_s}.\vc{q}}{m}
\pm \epsilon^B_{\vc{q}}$,
$\gamma^\pm_{\vc{q}} \simeq \pm
{\rm Im}{\cal K}_n(\vc{q},\omega^\pm_{\vc{q}})
c^2
\vc{q}^2/\epsilon^B_{\vc{q}}$.

The substitution of the solution of Eq.(\ref{rho}) into
(\ref{n}) allows finally to get the GRPA kinetic
equation for a Bose gas \cite{condenson}. 
As explained 
in the introduction we find that 
${\partial}n_{\vc{k_s}}/{\partial t}=0$ i.e. the condensed particle number
is a constant of motion
and thus no particle is exchanged  
with the normal fluid. For 
the normal fluid, the populations in each mode
obey
the balance equations:
\begin{eqnarray}\label{K6}
\frac{\partial}{\partial t}n'_{\vc{k}}=
\sum_{\vc{q},\vc{k'}} \left|
\frac{\frac{8\pi a}{mV}}
{{\cal K}(\vc{q},\epsilon_{\vc{k}+\vc{q}}-\epsilon_{\vc{k}})}
\right|^2 \!\!
\pi \delta(\epsilon_{\vc{k}+\vc{q}}+\epsilon_{\vc{k'}-\vc{q}}-
\epsilon_{\vc{k}}-\epsilon_{\vc{k'}})
\nonumber \\
\big[n'_{\vc{k}+\vc{q}}n'_{\vc{k'}-\vc{q}}
(n'_{\vc{k}}+1)(n'_{\vc{k'}}+1)-
n'_{\vc{k}}n'_{\vc{k'}}(n'_{\vc{k}+\vc{q}}+1)(n'_{\vc{k'}-\vc{q}}+1)\big]
\end{eqnarray}
These show that 
two normal particles 
collide through emission and absorption of a Bogoliubov collective 
excitation, called ``condenson'' (see Fig.1). The dispersion relation 
for an excitation of momentum $\vc{q}$ is given by 
the pole of the DDF
which plays the role of a propagator 
in (\ref{K6}). For a momentum $\vc{q}$, this excitation has 
the energy $\omega_\vc{q}$ and a decay rate, 
due to its absorption by a thermal atom, given by 
the Landau damping $\gamma_{\vc{q}}$.
Similarly to the plasma case\cite{Ichimaru}, these equations are valid 
provided that  
the collective oscillations
are damped i.e. 
$\gamma_{\vc{q}} \geq 0$. Otherwise, they 
are amplified
exponentially and the Bose gas becomes 
unstable.
The case with instability leads to different kinetic 
equations and is not treated here. 
But the instable regime 
allows particles exchange between the two fluids.
The equilibrium solution of (\ref{K6}) for a normal fluid 
at rest is given by $n'^{eq}_{\vc{k}}=
1/\exp{[\beta(\epsilon_\vc{k}-\mu)]}-1$. 
In this case when 
the temperature is close to zero, 
we find that:
\begin{eqnarray}
{\rm Im}{\cal K}_n(\vc{q},\omega)
=
\frac{2a m}{\beta |\vc{q}|}
\ln\left(\frac{1-\exp\{-\beta[
m\frac{(\omega+\vc{q}^2/2m)^2}{2\vc{q}^2}
-\mu]\}}
{1-\exp\{-\beta[
m\frac{(\omega-\vc{q}^2/2m)^2}{2\vc{q}^2}
-\mu]\}}\right)
\end{eqnarray}
which 
is a positive function for
$\omega=\omega^\pm_{\vc{q}} \geq 0$ and negative otherwise. Thus, 
we deduce that 
the stability condition $\gamma^\pm_{\vc{q}} \geq 0$ is equivalent to  the Landau criterion 
$\epsilon^B_{\vc{q}}\geq \vc{k_s}.\vc{q}/m$ i.e. $|\vc{k_s}/m| \leq c$.
In other words, the condensate is
stable if its velocity relative to the normal fluid is
smaller than the sound velocity.

\section{THE DYNAMIC STRUCTURE FACTOR}

The dynamic structure factor is defined as:
\begin{eqnarray}\label{S}
S(\vc{q},\omega)=\lim_{\tau \rightarrow \infty}
\int_V \!\!d\vc{r}\int_{-\infty}^\infty \! \frac{d\tau'}{2\pi}
e^{-i(\vc{q}.\vc{r}-\omega \tau')}
\langle \rho_{\vc{q}}(\tau)\rho_{-\vc{q}}(\tau+\tau')\rangle
\end{eqnarray}
This function is calculated within
the GRPA and
within the assumptions that 
the normal fluid is in thermal equilibrium at rest.
We start with the initial condition that for $\tau=0$ 
the correlation function contains only non interacting 
contribution\cite{condenson}:
$\langle \rho_{\vc{k'},-\vc{q}}
\rho_{\vc{k},\vc{q}} \rangle|_{\tau=0}=
(n_{\vc{k}}+1)n_{\vc{k'}}
\delta_{\vc{k'}-\vc{k},\vc{q}}(1-\delta_{\vc{q},0})
+n_{\vc{k}}n_{\vc{k'}}
\delta_{\vc{q},0}$.
Through Eq.(5), the evolution of the system will create 
for $\tau \rightarrow \infty$ 
some correlation due to the presence of the interaction 
between particles.
Similarly to the plasma case, 
the resulting dynamic structure factor is proportional to the inverse 
of the DDF \cite{Ichimaru}:
\begin{eqnarray}\label{S2}
S(\vc{q},\omega)=
\delta(\omega)\delta_{\vc{q},\vc{0}}
N^2
+\frac{1}{2\pi}
\frac{1}{\exp(-\beta \omega)-1}
\frac{mV}{4\pi a} {\rm Im}
\left( \frac{1}{{\cal{K}} (\vc{q},\omega)} \right)
(1 - \delta_{\vc{q},\vc{0}})
\end{eqnarray}
The relevant contributions are near the
poles
of ${\cal K}(\vc{q},\omega)$.
At temperature close to zero, 
we obtain the approximated expression:
\begin{eqnarray}\label{1/K}
\frac{1}{{\cal{K}} (\vc{q},\omega)}
=\frac{c^2 \vc{q}^2}{\epsilon^B_{\vc{q}}}
\left(\frac{1}{\omega -\omega^+_{\vc{q}}+i\gamma^+_{\vc{q}}}
+
\frac{1}
{\omega - \omega^-_{\vc{q}}+i\gamma^-_{\vc{q}}}\right)
\end{eqnarray}
Eq.(\ref{S2}) and (\ref{1/K}) are the main  
results of this paper and generalize previous results \cite{Ketterle,Stringari} 
for any temperature in the presence of a relative velocity between the 
two fluids. 
Due to the asymmetric 
dispersion relation,
this factor has two asymmetric peaks with 
usual intensity given by $S(\vc{q})=\vc{q}^2/2m \epsilon^B_\vc{q}$. 
Also the presence of temperature gives rise to the 
broadening of the peaks with a width corresponding 
to the Landau damping.  
When the relative velocity 
approaches the sound velocity,  
the asymmetric Bogoliubov frequency approaches zero 
for some non zero momentum for which the  Landau damping gets closer to zero.
Beyond that velocity, in some regions of the momentum, 
both the resonance frequency and the Landau damping become 
negative.
In that case, the gas becomes 
instable leading to non equilibrium dynamics.
Thus the experimental analysis of the dynamic structure factor 
provides a different way to confirm that the sound 
velocity is the critical velocity beyond which some 
instability should appear.
In real experiments, the width comes mainly from 
inhomogeneous broadening but, in principle, the asymmetry 
between the peaks must be observed. For this 
purpose, we need to create a situation with a non zero relative 
velocity\cite{Ketterle2}. Knowing the spatial density and velocity 
distribution of the condensate, we can use the local density 
approximation to calculate the line shape of the Bragg excitations 
and test the validity of the results \cite{Ketterle,Stringari}.

To conclude, the GRPA formalism confirms previous results established 
for the dynamic structure factor. Moreover, it allows 
to calculate this quantity from (\ref{S}) not only for 
an equilibrium distribution function but also for a 
non equilibrium one. Of special interest is the metastable 
state of superfluidity from which an explicit expression has been 
obtained for temperature close to zero and which  displays the transition 
from metastability  to instability. 
This transition is dictated by the Landau criterion which delimits 
the region in which the dynamic structure factor is well defined. 
Beyond that region this function is not stationary anymore due to  
unstable dynamics. 

\section*{ACKNOWLEDGMENTS}

PN thanks W. Ketterle, J. Tempere and J. Devreese for discussions.
PN acknowledges financial support from the Communaut\'e Fran\c caise de
Belgique under grant
ARC 00/05-251, from the IUAP programme of the Belgian
government under grant V-18, and from the EU under project RESQ
(IST-2001-35759).


\begin{thebibliography}{9}
\bibitem{condenson}
P. Navez, cond-mat/0309319.
\bibitem{Leggett}
A. J. Leggett, Rev. Mod. Phys. {\bf 73}, 307 (2001).
\bibitem{Zaremba} E. Zaremba, T. Nikuni, and A. Griffin,
J. Low Temp. Phys. {\bf 116}, 227 (1999).
\bibitem{Walser}  R. Walser, J. Williams, J. Cooper, M. Holland,
Phys. Rev. A, {\bf 59}, 3878 (1999).
\bibitem{Ketterle}
D.M. Stamper-Kurn, A.P. Chikatur, A. G\"orlitz, S. Inouye, S. Gupta, 
D.E. Pritchard, and W. Ketterle, Phys. Rev. Lett. {\bf 83}, 2876 (1999). 
\bibitem{NP} D. Pines and P. Nozi\`eres,
{\it The theory of Quantum Liquids}, vol 1 (Benjamin, New York,1966).
\bibitem{Ichimaru} S.Ichimaru, {\it
Frontier in Physics: Statistical Plasma Physics, Vol I: Basic Principle}
(Addison Wesley, 1992).
\bibitem{Stringari}
F. Zambelli, L. Pitaevskii, D.M. Stamper-Kurn and S. Stringari,
Phys. Rev. A {\bf 61}, 063608 (2000).
\bibitem{Ketterle2}
D.M. Stamper-Kurn, H.-J. Miesner, S. Inouye, M.R. Andrews, and W. Ketterle,
Phys. Rev. Lett. {\bf 81}, 500 (1998).
\end{thebibliography}
\end{document}